\documentclass[fleqn,10pt]{wlscirep}
\usepackage[utf8]{inputenc}
\usepackage[T1]{fontenc}
\usepackage{amsmath}
\usepackage{cite}
\usepackage{wrapfig}
\usepackage{multirow}
\usepackage{graphics} 
\usepackage{epsfig} 
\usepackage{graphicx}
\usepackage{caption}
\usepackage{subcaption}
\usepackage{multirow}
\usepackage{colortbl}

\graphicspath{{./figures/}}

\title{Development and Validation of Fully Automatic Deep Learning-Based Algorithms for Immunohistochemistry Reporting of Invasive Breast Ductal Carcinoma
}

\author[1]{Sumit Kumar Jha}
\author[1*]{Purnendu Mishra}
\author[1]{Akash Modi}
\author[1]{Gursewak Singh}
\author[1]{Shubham Mathur}
\author[1]{Rajiv Kumar}
\author[1]{Kiran Aatre}
\author[1]{Suraj Rengarajan}
\affil[1]{Applied Materials India Pvt. Ltd., Bangalore, India}

\affil[*]{purnendu\_mishra@amat.com (corresponding author)}

\begin{abstract}
Immunohistochemistry (IHC) analysis is a well-accepted and widely used method for molecular subtyping, a procedure for prognosis and targeted therapy of breast carcinoma, the most common type of tumor affecting women. There are four molecular biomarkers namely progesterone receptor (PR), estrogen receptor (ER), antigen Ki67, and human epidermal growth factor receptor 2 (HER2) whose assessment is needed under IHC procedure to decide prognosis as well as predictors of response to therapy. However, IHC scoring is based on subjective microscopic examination of tumor morphology and suffers from poor reproducibility, high subjectivity, and often incorrect scoring in low-score cases. In this paper, we present, a deep learning-based based semi-supervised trained, fully automatic, decision support system (DSS) for IHC scoring of invasive ductal carcinoma.  Our system automatically detects the tumor region removing artifacts and scores based on Allred standard. The system is developed using 3 million pathologist-annotated image patches of $512 \times 512$ from 300 slides, 50 thousand inhouse cell annotations, and 40 thousand pixels marking of HER2 membrane. We have conducted multicentric trials at four centers with three different types of digital scanners in terms of percentage agreement with doctors for Ki67 – 94\%, HER2 – 92\%, ER – 88\%, and PR - 82\%.  In addition to overall accuracy, we found that there are 5\% of cases where pathologist has changed their score in favor of automated system score while reviewing with detailed algorithmic analysis. Our approach could improve the accuracy of IHC scoring and subsequent therapy decisions, particularly where specialist expertise is unavailable. Our system is highly modular. The proposed algorithm modules can be used to develop DSS for other cancer types.

\end{abstract}

\begin{document}

\flushbottom
\maketitle
%
%
\thispagestyle{empty}


\section*{Introduction}
Breast cancer is a significant and escalating global health concern, with an alarming growth rate affecting millions of women worldwide. According to the World Health Organization (WHO), breast cancer is the most diagnosed cancer in women, accounting for approximately 25\% of all cancer cases.  In recent years, the incidence of breast cancer in women has exhibited an upward trend, emphasizing the critical need for robust diagnostic tools that can aid pathologists in accurately assessing IHC-stained tissue samples. According to the Global Cancer Observatory (GCO), the number of new breast cancer cases has witnessed a substantial increase, reaching over 2 million cases in 2020 alone. Moreover, projections indicate a continuous rise in breast cancer incidence in the coming years. These increasing cases are emphasizing the urgent need for effective diagnostic and treatment strategies.

Immunohistochemistry (IHC) staining plays a crucial role in determining the expression levels of key biomarkers, such as estrogen receptor (ER), progesterone receptor (PR), human epidermal growth factor receptor 2 (HER2), and Ki67. These biomarkers provide vital information for accurate diagnosis, prognosis, and treatment decisions in breast cancer patients. However, the manual interpretation of IHC stains is a laborious and time-consuming process, prone to inter-observer variability, exacerbated by the growing number of breast cancer cases and the shortage of pathologists available to handle the increasing workload. 

To address this mounting challenge, the collaboration between pathologists and advanced technologies becomes imperative. The integration of cutting-edge technology can significantly augment the diagnostic capabilities of pathologists, enabling accurate and efficient analysis of large volumes of IHC-stained tissue samples. In particular, the development of automated and objective scoring algorithms has the potential to revolutionize breast cancer diagnostics and mitigate the burden on pathologists.
The automated IHC scoring solution must detect the tumor region, and regions of interest (ROI), first and then compute stain-specific scores in the tumor region only. There are many semi-automated solutions where the user delineates the tumor region and runs a scoring algorithm to get an IHC score. This kind of approach limits the automation and puts a burden on pathologists to demarcate tumor regions [REF]. The major challenge is the time involved in marking the regions of interest within the tissue.  The region marking time depends upon the tissue size and the number of tumor regions within the tissue under investigation. Additionally, there are certain dependencies on the tool used to perform these markings. 

In our study, we propose a novel semantic segmentation-based training model for breast cancer IHC scoring [ ROI + SCORE]. Unlike transfer learning, which utilizes pre-trained models, our approach involves training a CNN from scratch specifically for IHC staining analysis. The model is trained using a large dataset of annotated IHC images, where the objective is to accurately segment the stained regions corresponding to the biomarkers of interest. By directly training the model for semantic segmentation, we aim to capture the fine-grained details and spatial relationships within the stained regions, leading to improved scoring accuracy and robustness.
In addition, we conducted a rigorous multi-centric trial to evaluate the performance and robustness of our semantic segmentation-based training model. The trial involved multiple institutions, laboratories, and pathologists, ensuring the algorithm's reliability and generalizability across different settings. The performance of our model was compared against manual scoring by experienced pathologists, serving as the ground truth for comparison.
The results of the multi-centric trial demonstrated the superiority of our semantic segmentation-based training model over conventional image processing techniques, SVM/RF-based approaches, and even transfer learning methods. Our model achieved higher accuracy, reduced inter-observer variability, and better adaptability to variations in staining patterns, slide preparation techniques, and scanner characteristics.

Our study proposes a novel semantic segmentation-based trained model for breast cancer IHC scoring, which outperforms algorithms based on conventional image processing techniques, SVM/RF-based approaches, and transfer learning methods. By training the model from scratch, we can capture the intricate staining patterns and spatial relationships, leading to improved accuracy and robustness. The findings of our multi-centric trial validate the effectiveness and reliability of our model, suggesting its potential for widespread clinical implementation in breast cancer diagnosis and treatment decision-making. 

This paper presents IHC breast cancer solutions in three sections. The first section discusses the automatic tumor region detection (Region of Interest) algorithm. The second section talks about nuclei stain-based algorithms related to ER, PR, and KI67 markers, and the third section presents membrane-based stain i.e., HER2 marker.

\section*{Results}
\subsection*{Evaluation Metrics}
\subsubsection*{Clinical Evaluation Metrics}

The ER and PR scores are calculated in terms of Allred scoring format~\cite{fauzi2022allred}, that is, score ranges are [0, 8] and it has three components Intensity Score (IS), Proportion Score (PS), and Total Score (TS). TS is the sum of IS and PS. While $IS \in [0,3]$, the $PS \in [0,5]$. The clinical evaluation of ER and PR scores is calculated by first converting the Allred score to categorical form given as

\begin{align}
    score (S) = \begin{cases}
negative, &\text{$TS < 3$}\\
positive, &\text{$TS \geq 3$}
\end{cases}
\end{align}

In the case of Ki67, the WSI score is given in terms of proliferation score (PRS) which ranges from [0,100]. The proliferation score is converted to the categorical form as given below.

\begin{align}
    score (S) = \begin{cases}
negative, &\text{$PRS < 15$}\\
positive, &\text{$PRS \geq 15$}
\end{cases}
\end{align}

The value of threshold score of 15 while deciding PRS in case of Ki67 is used commonly by pathologist in their diagnosis practice.

For, ER, PR, and Ki67 final performance is calculated in term of percentage agreement (PA) which is given as 

\begin{align}
    PA = \frac{1}{N} \left( \sum_{1}^{N}\chi(S_{GT_i},S_{PRED_i})\right) \cdot 100
\end{align}

where $\chi$ is an indicator function given as,

\begin{equation}
     \chi(S_{GT_i},S_{PRED_i}) = \begin{cases}
    1, & \text{if $S_{GT} = S_{PRED}$,}\\
    0, & \text{otherwise}
    \end{cases}  
\end{equation}
and $N$ is the number of samples.   

\subsection*{Performance}
\subsubsection*{Region of Interest (ROI)}
We have approximately 5 million patches extracted from 459 slides in training and 54 slides in validation distributed equally across stains, data sources, and scanners. The base model performance of 78\% Training accuracy and 73\% validation accuracy. Mean IOU across all channels is 0.70 for training and 0.64 for validation. The performance for final iteration in phase 2 was 80\% accuracy for Training and 75\% accuracy for validation. 55 slides kept aside for testing was validated on 2 classes (tumor and others) with final model from Phase 2. Our proposed model has achieved maximum IOU of 0.78 for tumor and 0.75 for others class.

\begin{table}[!t]
    \centering
    \resizebox{\linewidth}{!}{%
    \begin{tabular}{c|c|c|c|c}
     \hline
    & IOU & Precision vs Recall & Sensitivity vs Specificity & AUC \\ \hline
       Tumor vs Rest &  \begin{minipage}{.24\textwidth}
            \includegraphics[width=0.95\linewidth]{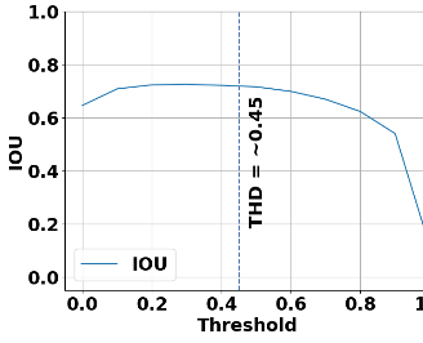}
            \end{minipage}  
          &
          \begin{minipage}{.24\textwidth}
            \includegraphics[width=0.95\linewidth]{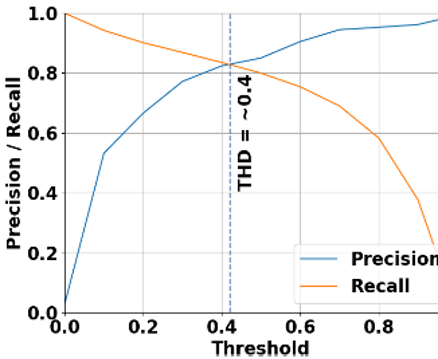}
            \end{minipage}  
           &
           \begin{minipage}{.24\textwidth}
            \includegraphics[width=0.95\linewidth]{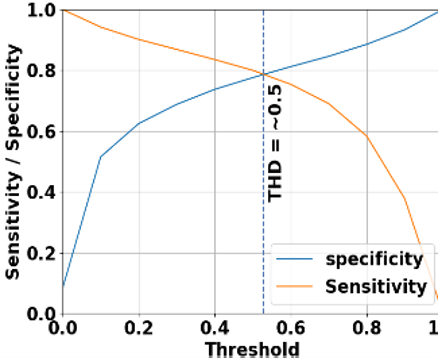}
            \end{minipage}
            &
            \begin{minipage}{.24\textwidth}
            \includegraphics[width=0.95\linewidth]{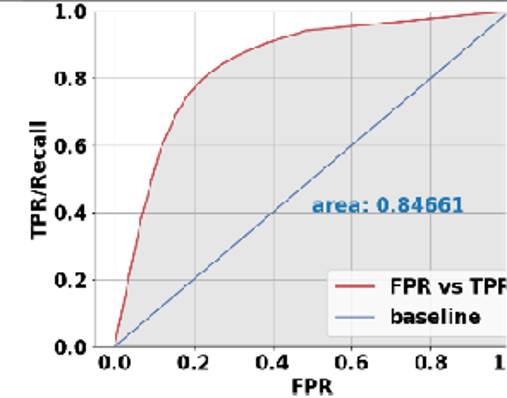}
            \end{minipage}  
            \\ \hline
      Other categories vs Rest   & \begin{minipage}{.24\textwidth}
            \includegraphics[width=0.95\linewidth]{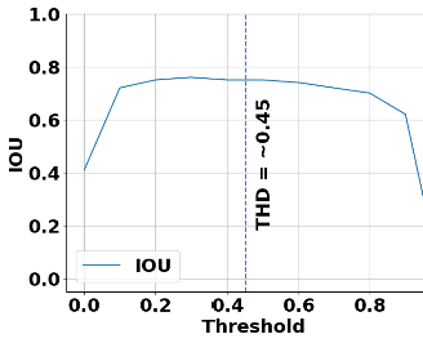}
            \end{minipage}
         & \begin{minipage}{.24\textwidth}
            \includegraphics[width=0.95\linewidth]{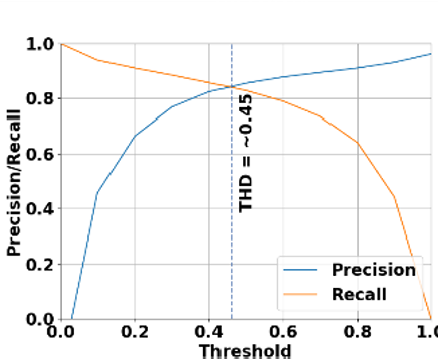}
            \end{minipage}
         & \begin{minipage}{.24\textwidth}
            \includegraphics[width=0.95\linewidth]{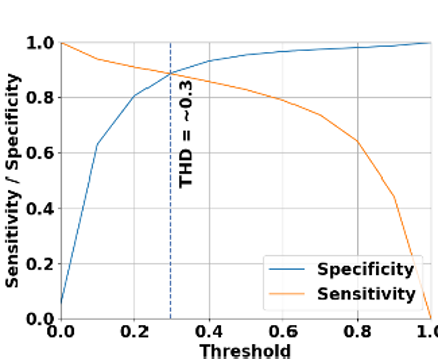}
            \end{minipage}
         & \begin{minipage}{.24\textwidth}
            \includegraphics[width=0.95\linewidth]{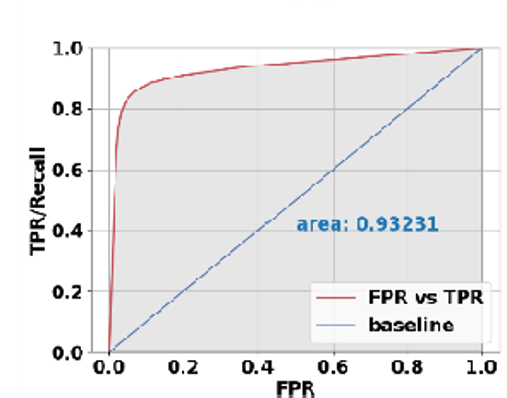}
            \end{minipage}  \\ \hline
    \end{tabular}%
    }
      \caption{Performance of final model. The first row has tumor vs everything else metrics. The second row has others versus everything else metrics. The values contained in columns left to right are IOU, precision, recall, sensitivity, specificity, and AUC Metrics, respectively}
    \label{tab:roi_performance}
\end{table}
  
The final model was also evaluated using various other metrics such as precision, recall, sensitivity, specificity, and AUC. The detailed performance of the model across all the metrics can be seen in Table~\ref{tab:roi_performance}. The optimum precision and recall value at intersecting threshold 0.4 is 0.83 for tumors and 0.81 at threshold 0.45 for others. Similarly, the optimum sensitivity and specificity value is 0.80 at 0.5 threshold for tumors and 0.90 at 0.3 threshold for others. All the plots show the model stability and robustness on the test set. The AUC is 0.85 for tumors and 0.92 for other classes. This tells that model predictions are ranked well, and the quality of the prediction is also good.

\subsubsection*{Nuclei segmentation and stain classification model’s performance evaluation}
Some sample nuclei segmentation and stain classification images are shown in Table~\ref{tab:nuclei_cells}. The top, middle, and bottom rows show tissue (left column) and algorithms prediction results (right column) samples for ER, PR, and Ki67, respectively. For Ki67, the stained nuclei are represented by red while blue represents the unstained tumorous nuclei. The light, moderate, and dark stained nuclei are represented with yellow, orange, and red for ER and PR. The blue colored line depicts the tumorous ROI identified by the ROI segmentation algorithm. Each detected nucle is represented by a mask with a black border.

\begin{table}[!t]
    \centering
    \begin{tabular}{c|c|c}
     \hline
    & Input & Nuclei Detection   \\ \hline
       ER &  \begin{minipage}{.33\textwidth}
            \includegraphics[width=0.95\linewidth]{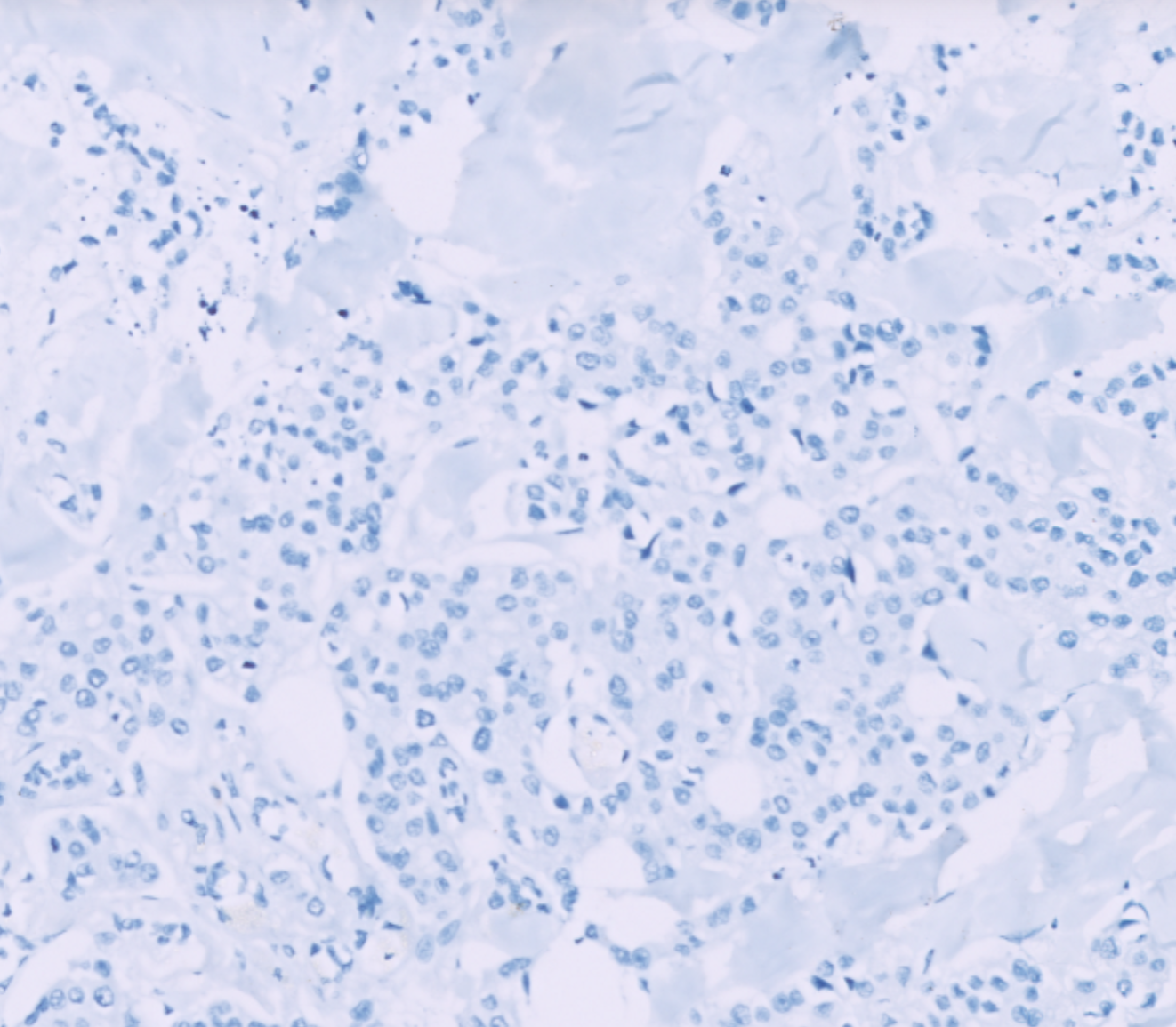}
            \end{minipage}  
          &
          \begin{minipage}{.33\textwidth}
            \includegraphics[width=0.95\linewidth]{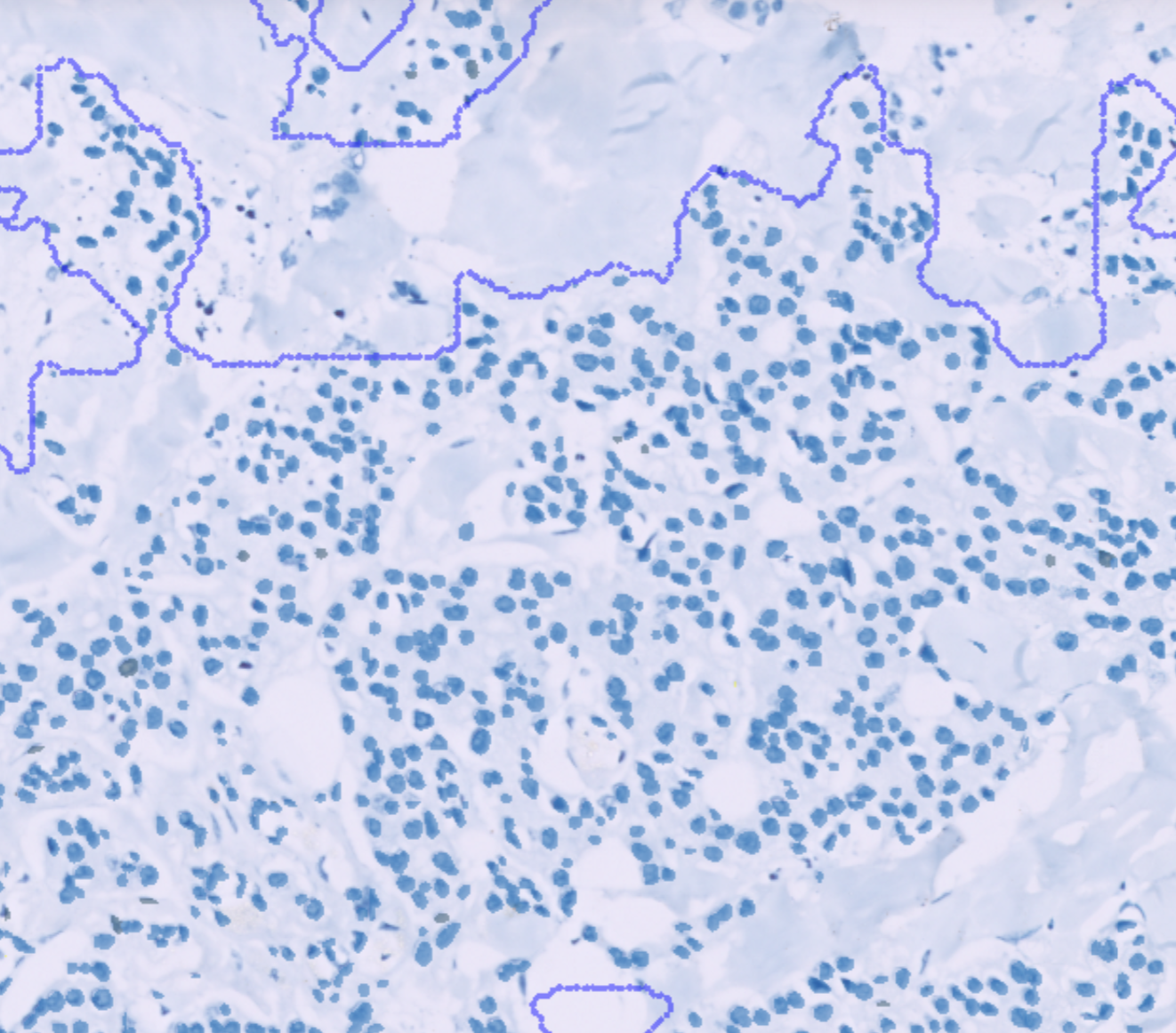}
            \end{minipage}  
            \\ \hline
        PR &  \begin{minipage}{.33\textwidth}
            \includegraphics[width=0.95\linewidth]{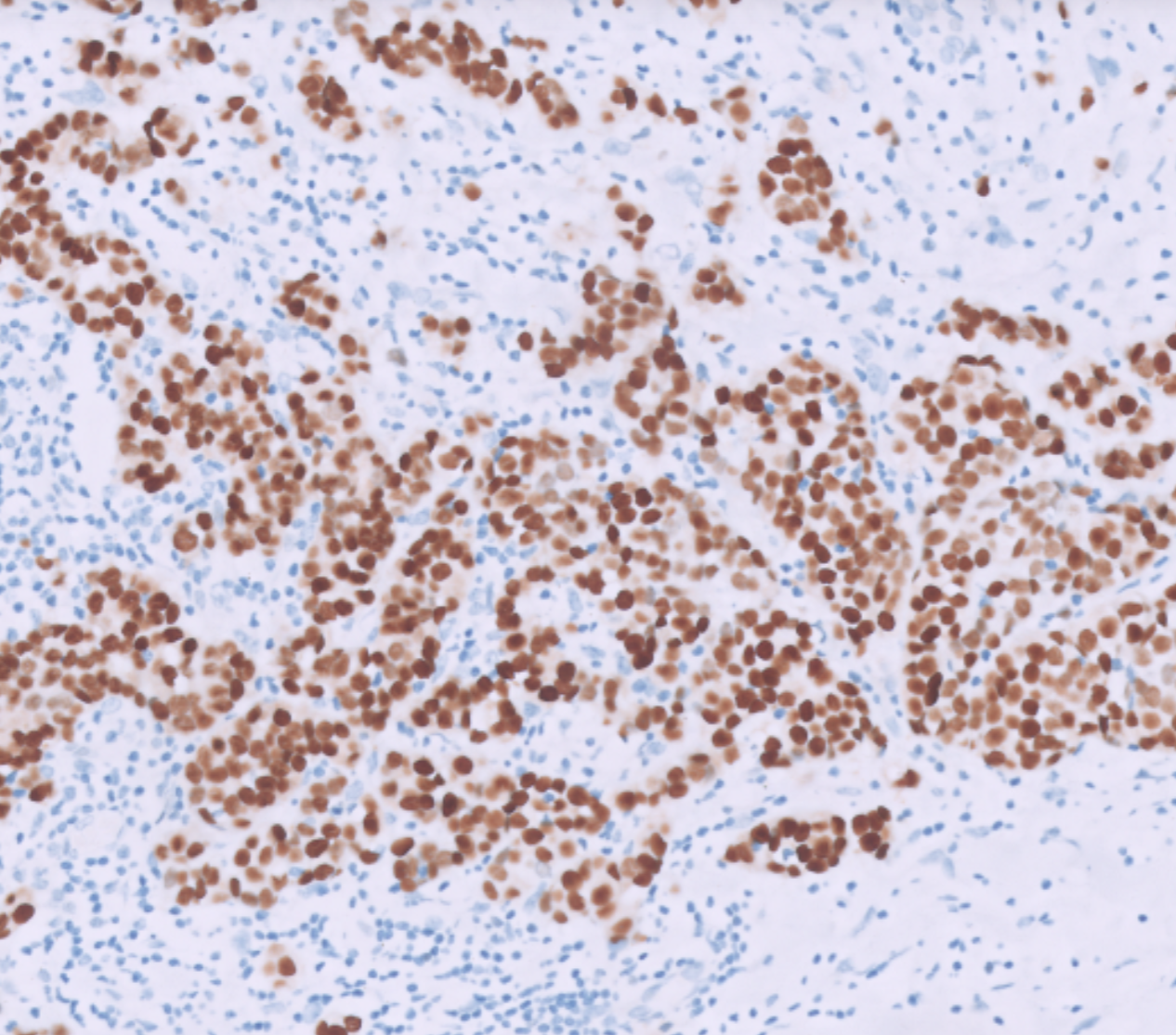}
            \end{minipage}  
          &
          \begin{minipage}{.33\textwidth}
            \includegraphics[width=0.95\linewidth]{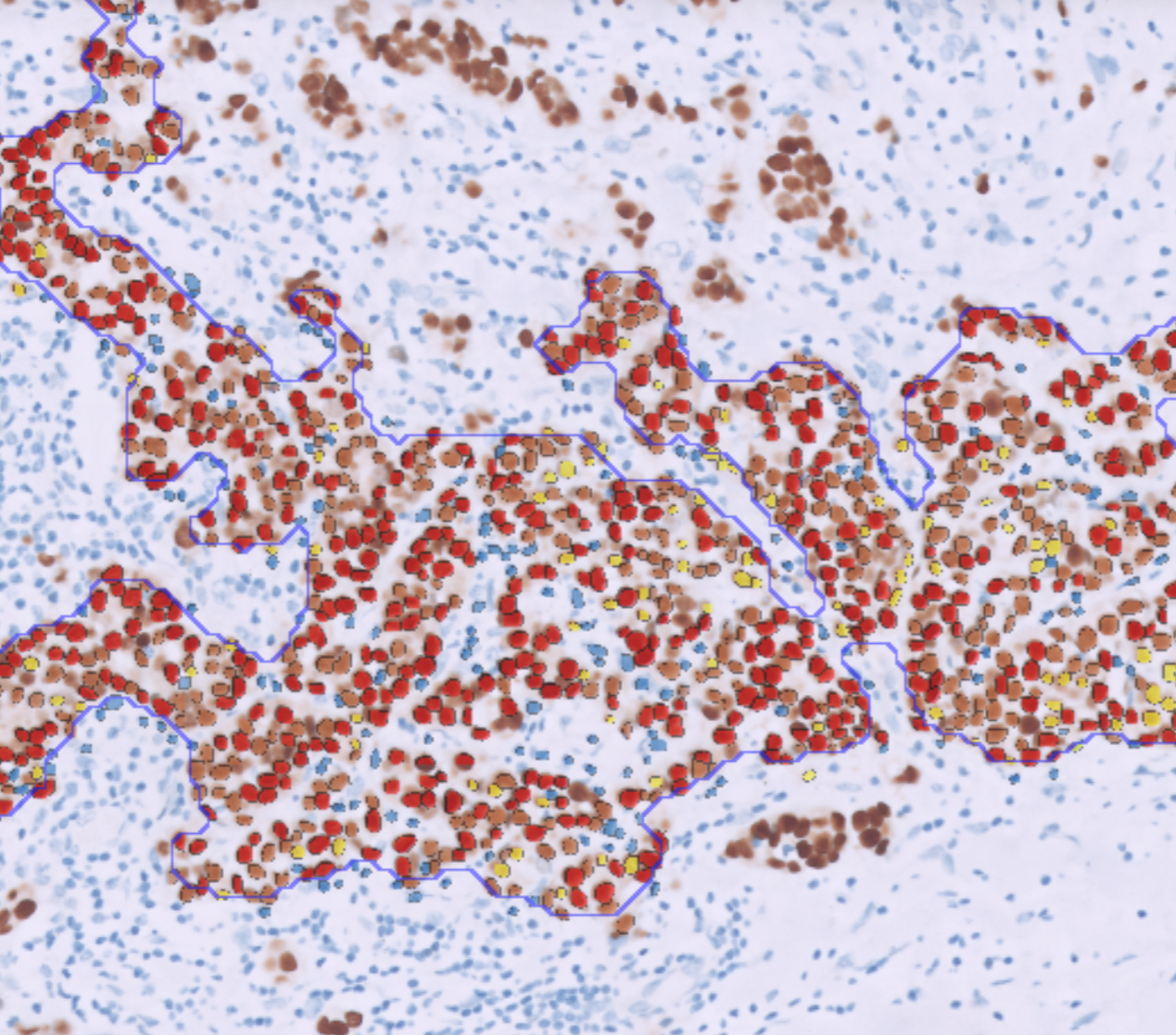}
            \end{minipage}  
            \\ \hline
         Ki67 &  \begin{minipage}{.33\textwidth}
            \includegraphics[width=0.95\linewidth]{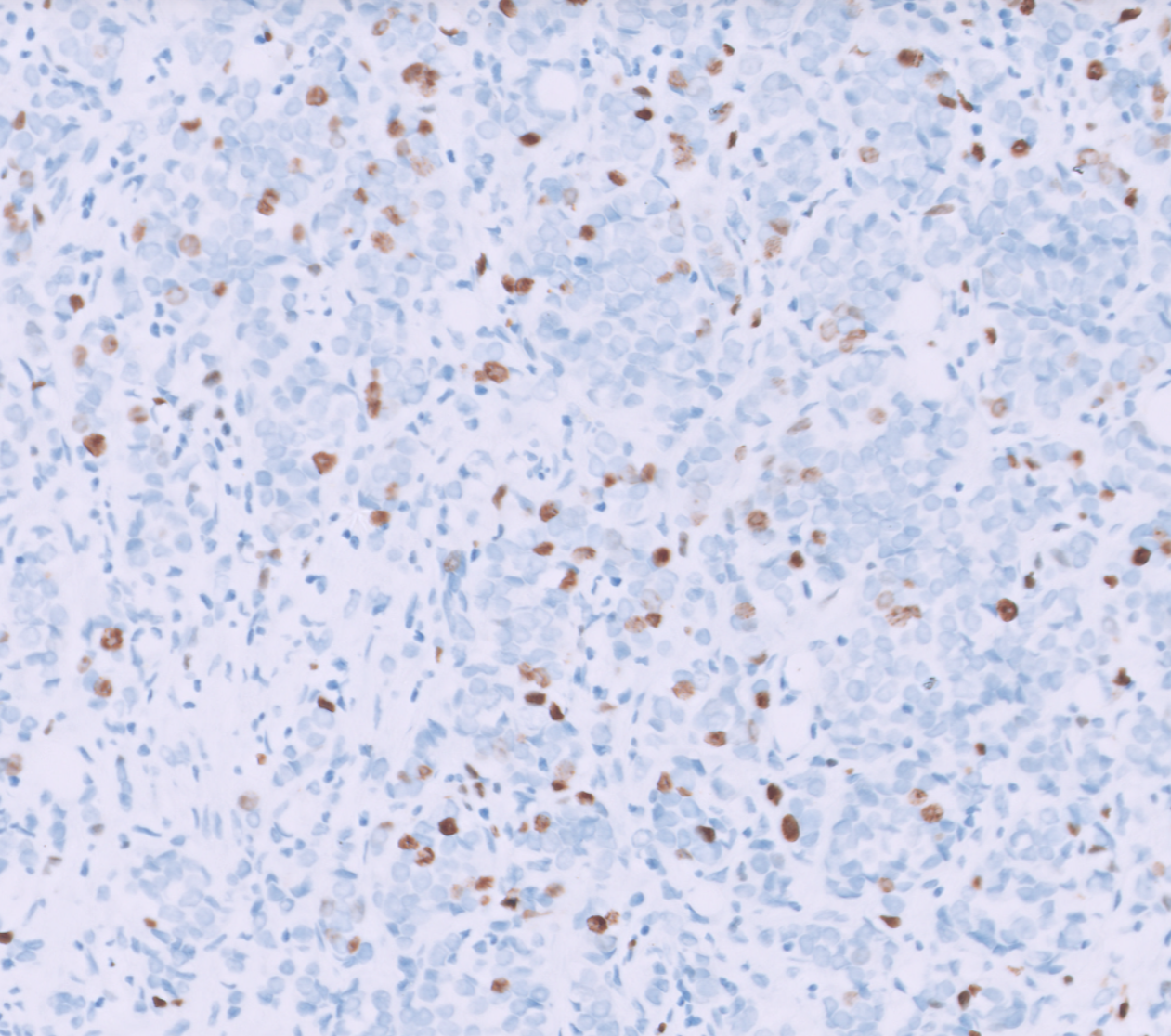}
            \end{minipage}  
          &
          \begin{minipage}{.33\textwidth}
            \includegraphics[width=0.95\linewidth]{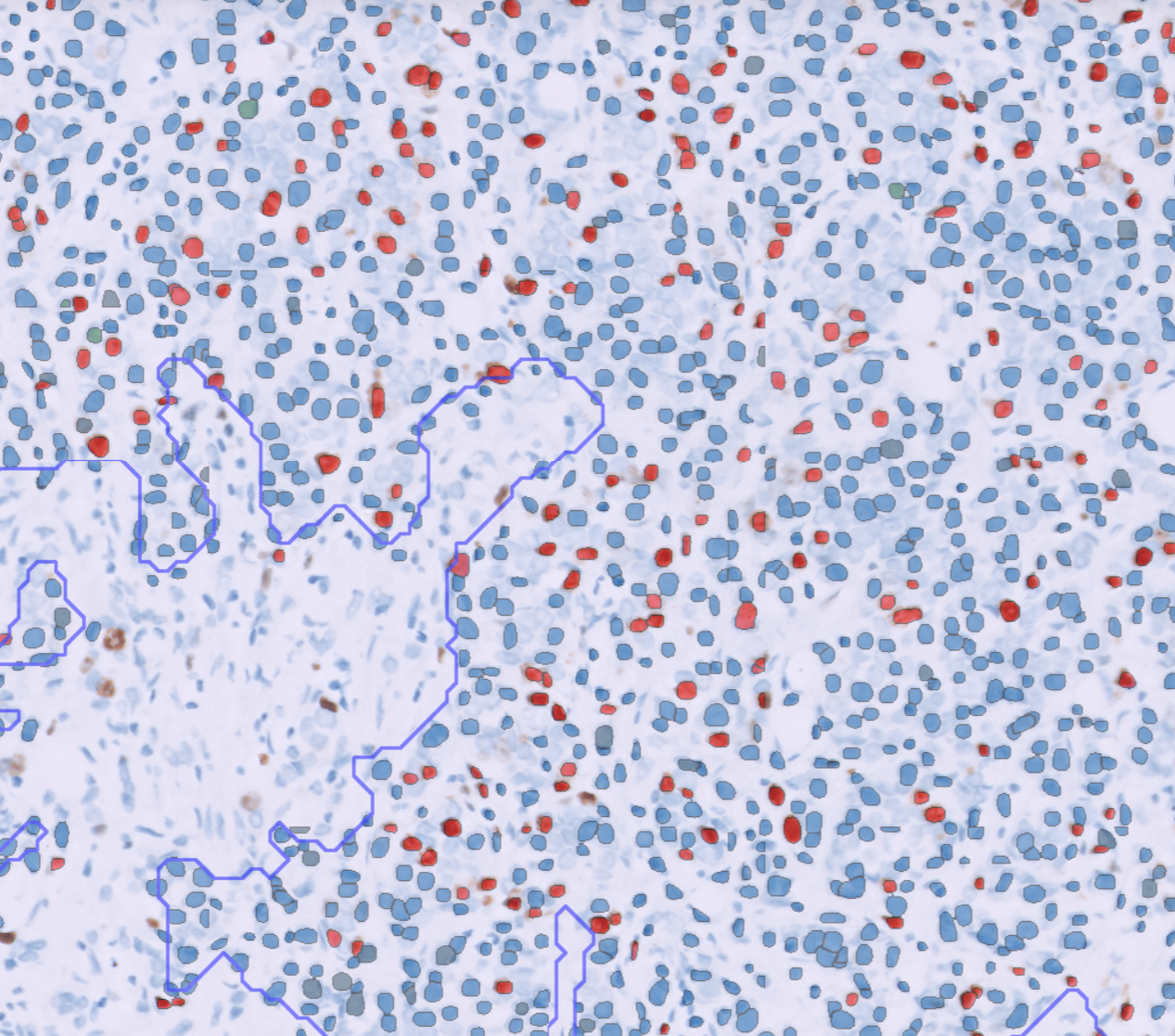}
            \end{minipage}  
            \\ \hline
    \end{tabular}
      \caption{Snapshots showing nuclei detection and stain classification results on samples taken from validation dataset for ER, PR, and Ki67 stained slides shown in top, middle, and bottom rows, respectively. (a) The snapshot of random region from a WSI, and (b) Overlayed nuclei detection and stain classification mask.}
    \label{tab:nuclei_cells}
\end{table}

The nuclei detection performance of the customized Mask-RCNN~\cite{he2017mask} model has been validated on a set of RGB patches of spatial dimensions $512 \times 512$. The patches are manually annotated to mark tumorous nuclei in the given for ground-truth creation. On the same patches, the prediction about the nuclei position is made using the trained model. The obtained prediction is compared with the ground-truth images. The comparison results are reported in Table~\ref{tab:nuclei_detection_performance}. The predicted cell count is very close to the number of nuclei marked by pathologists. The accuracy and $f_1{score}s$ are in a range of 75 – 85\%. This indicates the satisfactory performance of the model.

\begin{table}[!t]
\centering
\begin{tabular}{l|c|c|c|c|c|c}
\hline
Stain type & \multicolumn{1}{l}{GT cell count} & \multicolumn{1}{l}{Predicted cell count} & \multicolumn{1}{l}{Accuracy} & \multicolumn{1}{l}{Precision} & \multicolumn{1}{l}{Recall} & \multicolumn{1}{l}{F1score} \\ \hline
ER         & 13,509                            & 11,418                                   & 0.7884                       & 0.8344                        & 0.8616                     & 0.8478                      \\ \hline
PR         & 6,005                             & 5,378                                    & 0.7716                       & 0.7370                        & 0.8576                     & 0.7927                      \\ \hline
Ki67       & 3,243                             & 3,025                                    & 0.7534                       & 0.7357                        & 0.7983                     & 0.7657                      \\ \hline
\end{tabular}%
\caption{The performance of the nuclei segmentation model on a validation set of RGB patches.}
\label{tab:nuclei_detection_performance}
\end{table}

\begin{figure}[!t]
    \centering
    \subfloat[DNN method]{\includegraphics[width=0.33\linewidth]{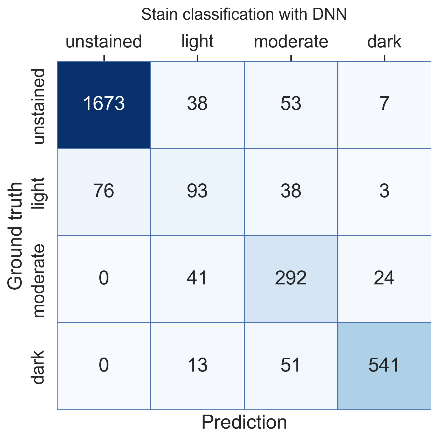}} 
    \subfloat[CMYK method]{\includegraphics[width=0.33\linewidth]{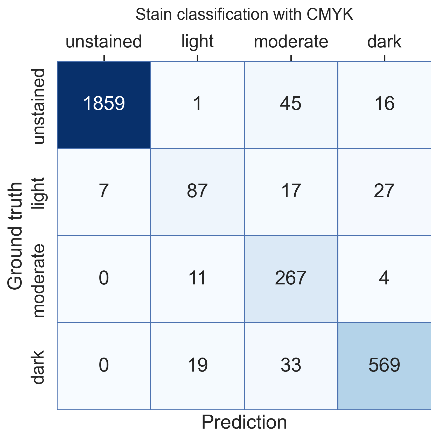}}
    \caption{The stain classification model’s stain classification performance of four different stain categories. (a) The performance of a DNN model (approach A1), and (b) the performance of CMYK (approach A2) based stain classification method.}
    \label{fig:cm_stain_classification}
\end{figure}

Similarly, the performance of two models (approaches) used for stain classification is calculated on a set of manually annotated ground-truth data by the field experts. The results are shown using confusion matrices in Figure~\ref{fig:cm_stain_classification}. The left matrix shows the performance of approach A1 (DNN-based method) and the right matrix shows the performance of approach A2 (CMYK-based method). The accuracies are 88.31\% and 93.92\% for A1 and A2, respectively. The major issue observed with the based approach is the misclassification of nuclei in adjacent stained cell categories. For example, a few times unstained cells were being labeled as light stained and vice-versa. A similar, issue was between light \& moderately stained, and moderately \& dark stained categories. The inter-class categorization issue was relatively lower in the CMYK-based approach (A2). Hence, this method is preferred over the DNN approach. Moreover, the inference time per nuclei was lower in CMYK based approach as compared to the DNN model. Thus, approach A2 enabled comparatively faster performance of the overall scoring pipeline.

\subsubsection*{Clinical Evaluation}
\begin{figure}[!t]
    \centering
    \subfloat[ER]{\includegraphics[width=0.33\linewidth]{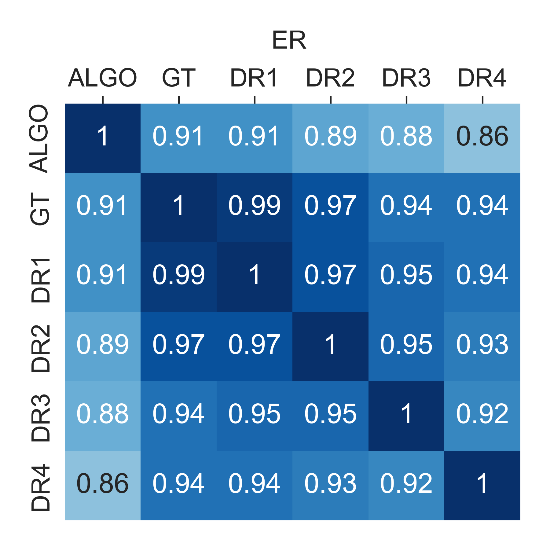}} 
    \subfloat[PR]{\includegraphics[width=0.33\linewidth]{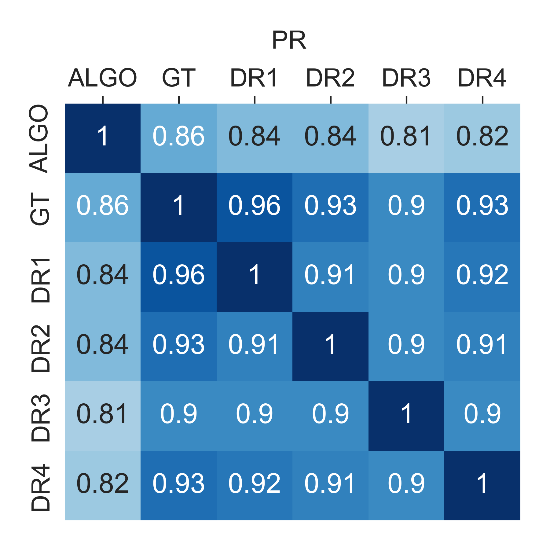}}
    \subfloat[Ki67]{\includegraphics[width=0.33\linewidth]{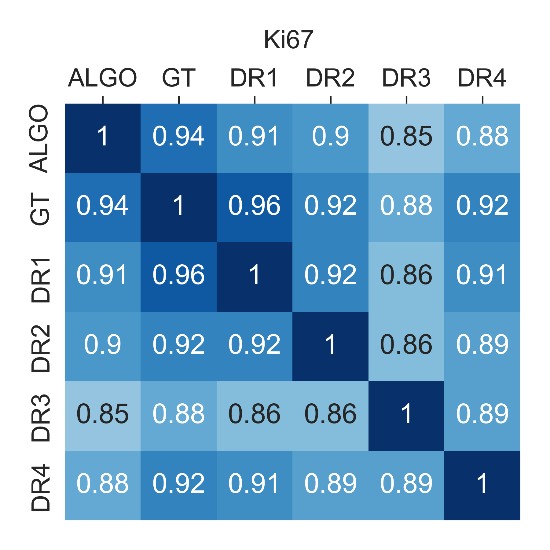}}
    \caption{Intra-observer results comparison on validation slides for (a) ER stain, (b) PR stain, and (c) Ki67 stain.}
    \label{fig:nuclei_clinical}
\end{figure}

The clinical evaluation of the deep learning-based ER, PR, and KI67 IHC marker used invasive ductal breast carcinoma grading is presented in the form of a confusion matrix. The confusion matrices for ER, PR, and Ki67 are shown in Figure~\ref{fig:nuclei_clinical}  Here, GT (ground truth) is the consensus score agreed by all four doctors. The confusion matrices show the algorithms' percentage agreement with GT as well as different pathologists. Additionally, the percentage of agreement between two different pathologists is shown. The results shown in the confusion matrices are for 421, 410, and 375 slides from ER, PR, and Ki67, respectively. These are the number of slides on which the scores from all four pathologists used for performance analysis were available. The ER and PR algorithms have approximately 91\% and 86\% agreement with consensus scores, respectively. For the ER stain type, the highest agreement value between two pathologists is 97\% while the lowest agreement value is 92\%. Similarly, for PR stain type, the highest and lowest agreement values between two pathologists are 92\% and 90\%, respectively. The algorithm achieved 94\% agreement with GT on Ki67 validation slides. While one of the pathologists 

The category-wise performance (clinical performance) of the algorithm measured in terms of percentage agreement with consensus score (GT) for ER, PR, and Ki67 on validation slides is furnished in Table~\ref{tab:nuclie_clinical}. The slide composition available for each category is also provided in Table X. The highest agreement of 94\% on 469 slides was achieved for Ki67 while the least agreement of 82\% on 510 slides for PR was achieved with an AI-based algorithm. The agreed baseline of the algorithm’s performance on validation slides was 85 +/- 5\%. The nuclei segmentation and classification algorithm has been able to meet this performance baseline.	

\begin{table}[!t]
\centering
\begin{tabular}{l|c|c|c}
\hline
         & ER (\#cases) & PR (\#cases) & Ki67 (\#cases) \\ \hline
Negative & 73 (178)     & 72 (230)     & 63 (65)        \\ \hline
Positive & 97 (339)     & 91 (280)     & 99 (414)       \\ \hline
Overall  & 89 (517)     & 82 (510)     & 94 (479)       \\ \hline
\end{tabular}
\caption{Category-wise percentage agreement of Algorithm with consensus ground-truth for ER, PR and Ki67 stained WSI.}
\label{tab:nuclie_clinical}
\end{table}

\section*{Discussion}
\begin{figure}[!t]
    \centering
     \subfloat[Sample patch affected by artifacts \label{subfig:input_patch}]{\includegraphics[width=0.47\linewidth]{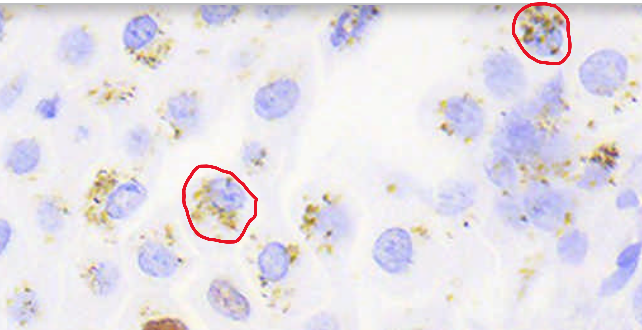}} \hfill
     \subfloat[Misclassification of some unstained cells as lightly stained \label{subfig:cell_misclassific}]{\includegraphics[width=0.47\linewidth]{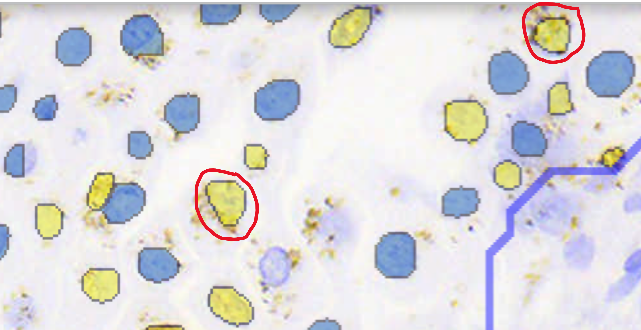}} 
    \caption{Highlighting the effect of artifacts on tissue on nuclei segmentation and stain classification. (a) A snapshot of the tissue region from ER WSI. (b) The nuclei segmentation and stain classification results.}
    \label{fig:nuclei_defect}
\end{figure}

The deep learning-based algorithm to automatically estimate the Allred and proliferation score for ER/PR and Ki67 stained tissue WSI works as envisioned, respectively. The whole processing pipeline combined with ROI model can segment tumorous nuclei and classify the nuclei stain color with acceptable accuracy. The scoring pipeline is analogous to the process followed in microscopic analysis of breast tissue. As the pathologist first select the tumorous regions within the tissue and then they grade nuclei within those regions give their score. In similar manner the AI-based algorithm segments out the tumorous region and another following algorithm generates a score based on detected nuclei count and their stain colors. However, the performance of the algorithms is not always as expected. It is affected by presence of defects, artifacts, acini, folds, image noises etc. in the breast tissue. The influence of these objects on overall for high score category of slides scores is comparatively less. However, they affect negative category slides more. This point is evident from the category-wise results reported in Figure 5. The performance of algorithm is slightly below exception on the negative category for all of three given stain types. An example illustrating the effect of artifact on stain classification process is shown in Figure~\ref{fig:nuclei_defect}. 

In Figure~\ref{subfig:input_patch}, a region of ER-stained tissue sample is shown. The shown region has artifacts around the tumorous nuclei. Two such nuclei are highlighted in red circle. While both nuclei are unstained, however due to the presence of artifacts the algorithm is predicting them as light-stained cells as shown in Figure~\ref{subfig:cell_misclassific}. The presence of artifacts and other similar defects in tissue leads the algorithm to provide higher scores for the tissue. This leads to degraded performance in low-score tissue slides. These types of scoring issues have been solved by introducing a post-processing module. One such approach is using clustering the tumorous nuclei based on their spatial location in tissue. It has been observed that the stained tissue generally occurs in a group of 6 or more nuclei while the false positive detected stained cells are likely to form smaller clusters. Thus, most of such small clusters of nuclei have been discarded during the score computations leading to better performance.
Another important point to highlight is that the algorithm analyzes all tissue tumorous regions with equal weightage. It considers very small regions in very large tissue segments which are sometimes missed by pathologists in the microscopic process leading to incorrect reporting. Several such instances were encountered in our study with pathologists involved with this project. The AI-based diagnostic solution has the ability to assist pathologists in well-informed and evidence-based reporting as well as improve the overall turn-around time for better patient outcomes.

\section*{Methodology}
\subsection*{Acquisition of data}
De-identified, digitized whole-slide images of IHC breast tissue for ER, PR, HER2, and Ki67 were obtained from multiple hospitals/laboratories. The slides were collected with the objective of the development and validation of automatic AI-based DSS for IHC breast tissue analysis. The study was proposed to be conducted in 2-phased - 1) Development and learning phase, and 2) multi-centric validation study phase. The development and learning phase were conducted in collaboration with Kasturba Medical College (KMC), Manipal. This phase involved the collection of retrospective samples that have previously been prepared by KMC, Manipal (KMC, MPL) and are authorized to be used for research purposes. In this phase, approximately 238 slides each of ER, PR, HER2, and Ki67 has been de-identified. The phase-1 study, and use of slides has been approved by the Institutional Ethics Committee. All the samples collected were retrospective samples that have been banked with KMC, Manipal. In order to ensure the good quality of the slides, the samples were restricted to being no older than 12 months.  The selected samples were anonymized and converted to digital TIFF images using a Morphle (Optimus 6T) scanner. Of the 952 slides (ER, PR, HER2, Ki67 excluding H\&E), 32 slides were rejected due to poor quality slide preparation, poor stain, folds tears etc. The total slides available for phase-1 is 920.

The phase-02 of study is the extension of phase-01 where the developed DSS has been validated on de-identified slides from 4 different centers were obtained.  In this study about 569 cases (555 ER, 559 PR, 555 HER2, and 554 Ki67)\footnote{For some of the cases out of 569, a full set of all four stains (ER, PR, Ki67, and HER2) was not received.}  were targeted. The institutes involved in this study were KMC (Manipal), KMC (Mangalore), Sikkim Manipal University (SMU), and Neuberg Anand Labs (NAALM). The approval for use of the slides has been taken from each individual Institute’s Ethical committee involved in this study. The details of the number of samples received from each of the institutes in phase-02 of the study have been provided in Table~\ref{tab:case_distribution} below. Out of the total 663 cases received, 94 cases were discarded due to bad quality of slides which includes reasons like out of focus, extreme folding of tissue \& water bubbles. All samples were duly anonymized in accordance with best practices. 

\begin{table}[!t]
\centering
\begin{tabular}{l|c}
\hline
\textbf{Institute}      & \textbf{Number of Cases} \\ \hline
KMC, Manipal   & 313             \\ \hline
KMC, Mangalore & 200             \\ \hline
SMU            & 50              \\ \hline
NAALAM         & 100             \\ \hline
\end{tabular}
\caption{Institute-wise cases distribution }
\label{tab:case_distribution}
\end{table}

The scanners utilized for the digitization of the slides were:
\begin{itemize}
    \item Morphle: Morpholens 6
    \item Morphle: Morpholens 240.
    \item Philips: Intellisite UltraFast Scanner UFS 300.
    \item Motic: Easyscan One
\end{itemize}

\subsection*{Collection of slide-level IHC scores}
We have given breast cancer cases (each case 4 stains - ER, PR, HER2, KI67) data to a panel of 5 pathologists to give the IHC scores for each case/patient. They have given their individual scores under the blind study condition. Later, consensus scores of all pathologists were collected on these cases to generate the ground-truth scores. Figure 6 illustrates the process of collection of slide-level scores from a group of pathologists. The consensus score collection became important when there was a major variability of scores observed between pathologists’ scores. The pathologists reviewed the slide through microscopes to provide IHC scores. They provided Allred scores for ER, and PR which include intensity scores (IS), proportion scores (PS), and total scores. For Ki67, we received proliferation scores in the range of 0 and 100. The IHC scores HER2 were reported from 0 to 3+ which is the measure of HER2 receptor protein on the surface of cells in a breast cancer tissue sample. All the slides were reviewed in a manner consistent with CAP guidelines with no time constraints.

\begin{figure}[!t]
    \centering
    \includegraphics[scale = 0.4]{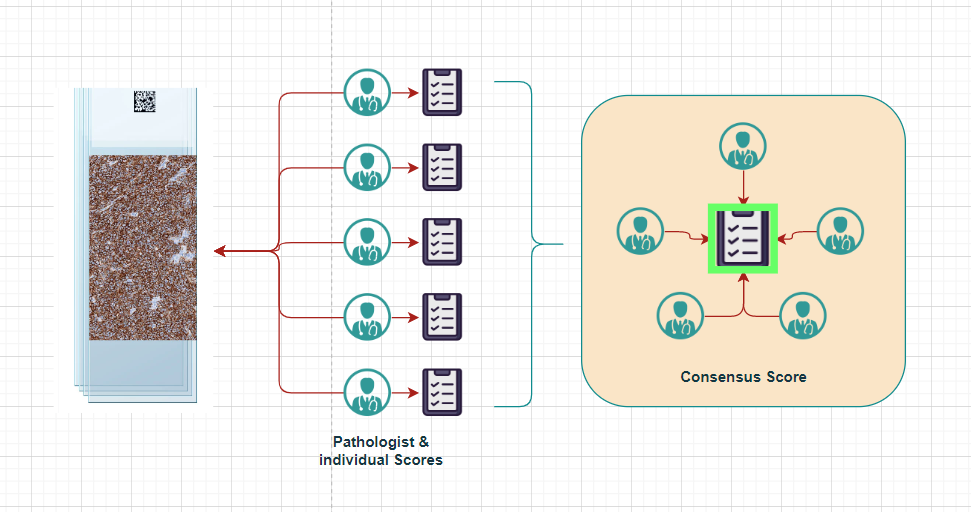}
    \caption{Illustration of the ground-truth generation process.}
    \label{fig:gt_generation}
\end{figure}

\subsection*{Collection of region-level annotations}
To train and validate the DL model’s region-level annotations were collected on the slides sample received during the phase-01 of the study. The pathologists and domain experts annotated tumorous and non-tumorous regions in the breast tissue sample. As there were multiple modules involved in the complete DSS, region-level annotations to meet the specific requirements of each of module is obtained. For example, the ROI module needs marking of tumorous, non-tumorous, artifacts, folds region within the breast tissue as part of training data source. Similarly, the ER and PR scoring system needed nucleus level information to train the model. Moreover, the HER2 module needed membrane markings to develop a scoring DL model. A brief overview of each of these annotation types and its collection process is provided below.  

\begin{figure}[!t]
    \centering
     \subfloat[Sample A]{\includegraphics[width=0.47\linewidth]{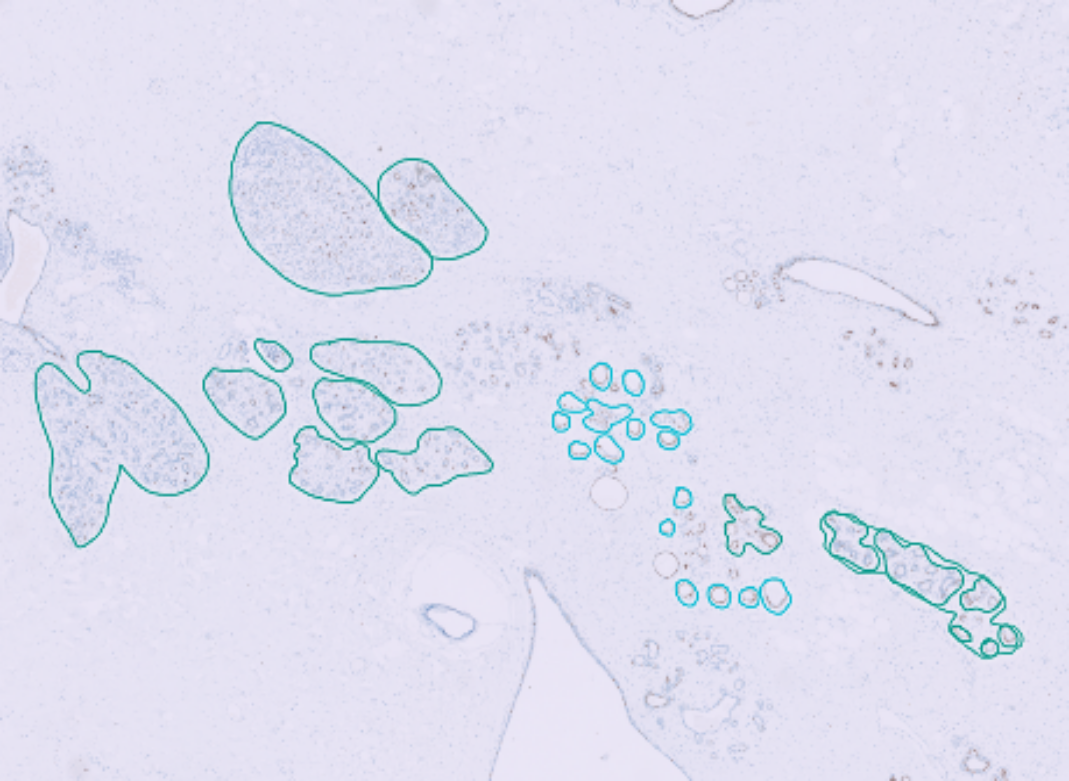}} \hfill
     \subfloat[Sample B]{\includegraphics[width=0.47\linewidth]{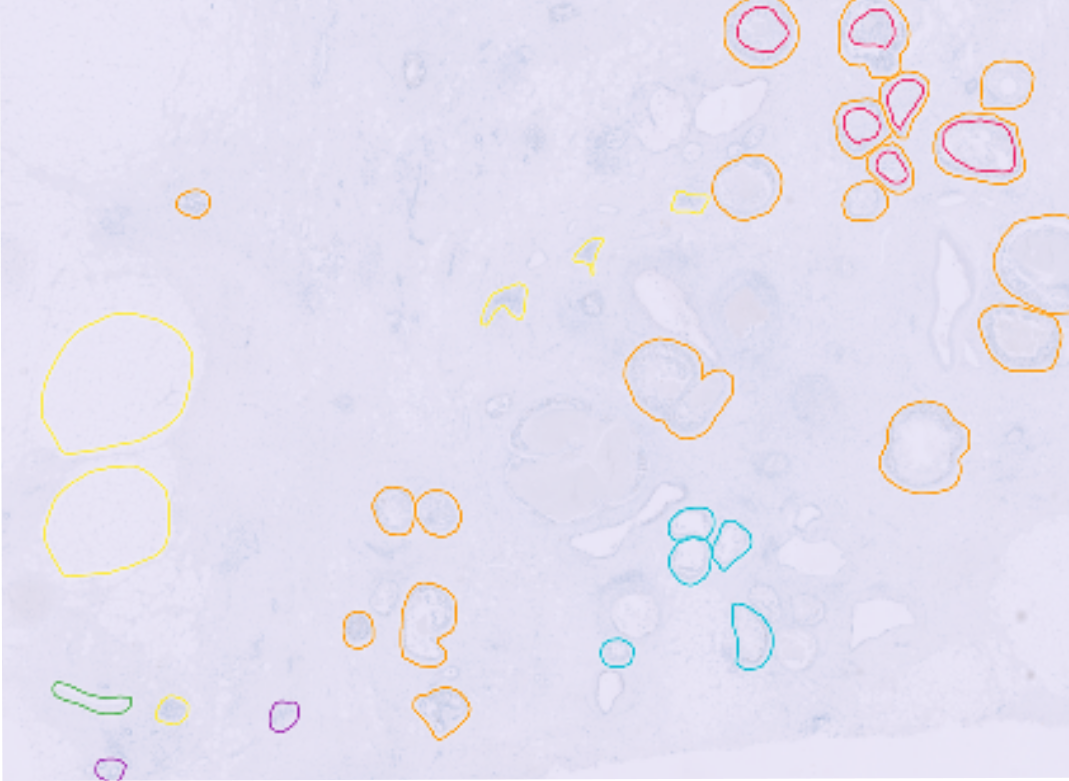}} 
    \caption{Examples of tissue sub-region annotations to be used for ROI deep learning model training and validation.}
    \label{fig:roi_marking}
\end{figure}

To obtain all possible and available variations in terms of slide color, stain process, different types of tissue regions, image resolution, etc. WSI was set to select from the available data bank and set aside for annotations. Since marking each individual region in the tissue is very difficult and time-consuming, regions within tissue referred to as super ROI are identified. These super ROIs are usually mutually exclusive and non-similar regions. Annotations within these are obtained for tumor, normal, stroma, acini, ducts, blood vessels, DCIS, etc. The regions like folds, artifacts, bubbles, of focus were also marked in the super ROI of tissue. All the regions were delineated using the free hand annotation tool available on the Mimansa AI web page. Figure~\ref{fig:roi_marking} shows examples of different regions marked by the domain experts to identify tumorous as well as non-tumorous within the tissue.

\begin{figure}[!t]
    \centering
     \subfloat[Sample A]{\includegraphics[width=0.47\linewidth]{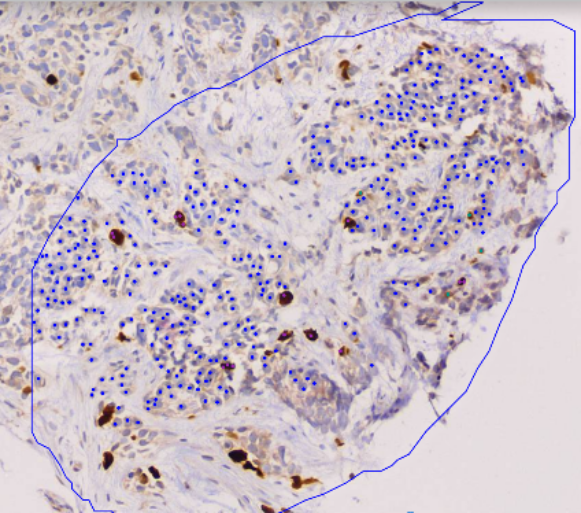}} \hfill
     \subfloat[Sample B]{\includegraphics[width=0.47\linewidth]{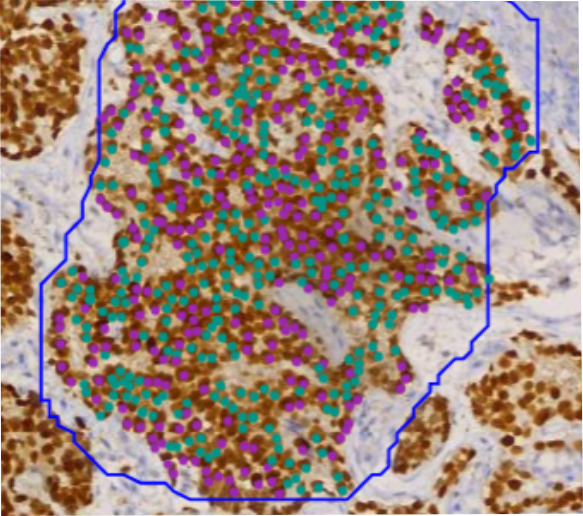}} 
    \caption{Example of tumorous nuclei annotation by domain experts.}
    \label{fig:nuclei_marking}
\end{figure}

\subsection*{Deep Learning System}
The decision support system (DSS) for breast tissue IHC analysis consists of two stages. The stage 01 corresponds to region of interest (ROI) detection while stage 02 is composed of two separate modules. One of the modules is for analysis and scoring of nuclei-based tissue analysis (ER, PR, and Ki67) while the second module is for membrane-based tissue analysis. All the models in both the stages are deep learning stage architectures. The output of stage 01 (ROI detection) is used in stage 02 for the specific tissue analysis given the tissue type. The details of each of the model is provided in the following sub-sections.

\subsubsection*{Region of Interest (ROI)}
The objective of the ROI detection model is to perform an automatic segment tumorous region for the breast tissue image. The task can be a semantic segmentation where each pixel in the input tissue is assigned a specific label. The based encoder-decoder UNET [29] architecture was tried to perform ROI segmentation. Though doing better on tumor regions, the results were not satisfactory, especially in terms of False Positives. Most of the Normal tissues such as acini and ducts were classified as tumors because they can also take up staining properties. There are specific tissues such as DCIS, squamous epithelium, etc. which are non-invasive carcinoma but look exactly like with \& without staining. These are very similar to invasive cancerous cells, but they are not tumorous yet. These regions were also getting segmented as tumors instead of discarding them. The improper/bad regions such as bubbles, artifacts, etc. were further adding to false positives due to their morphological similarity with cell structures.   

\begin{figure}[!t]
    \centering
    \includegraphics[width=\linewidth]{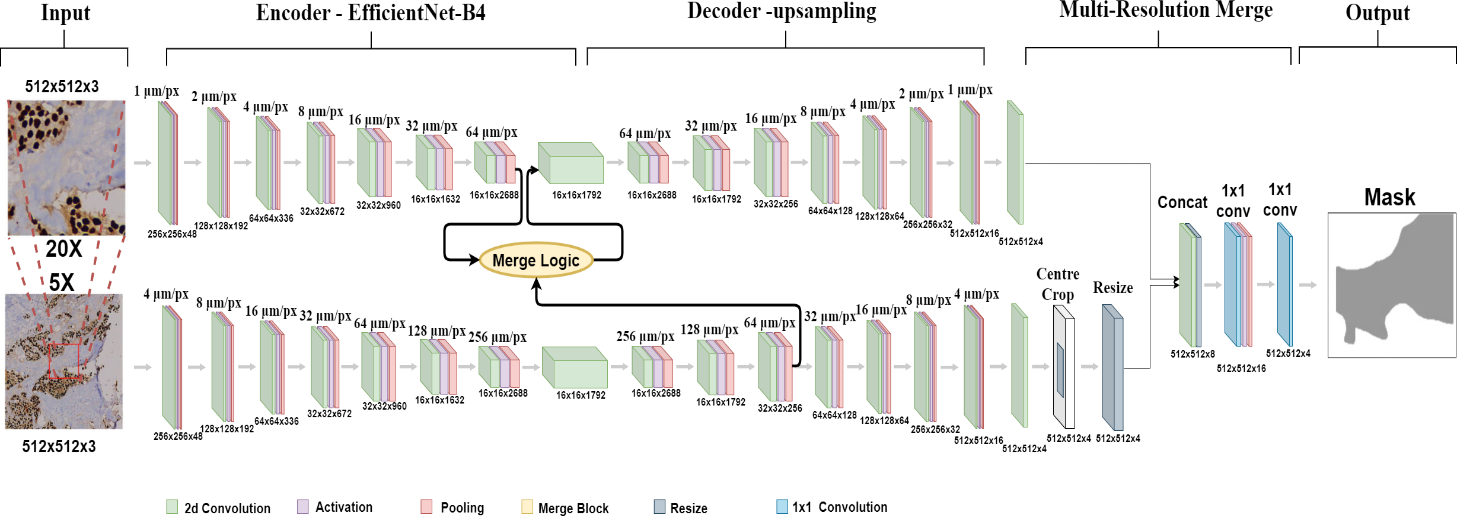}
    \caption{The architecture of multi-class multi-level UNet architecture with unique merge-logic and final concatenation layer to generate a single mask.}
    \label{fig:roi_model}
\end{figure}

A generic model is designed that should work with all four stain types that are ER, PR, Ki67, and HER2. Additionally, the stain-specific variation arising due to data coming from different laboratories and scanners should have a negligible effect on the model’s performance. Moreover, the model should multi-region classification from the given input tissue. But it adds up to certain segmentation complexities. For example, tumor region needs a local high-resolution field of view (FOV) whereas detection of other classes such as acini, Ducts, DCIS, etc. needs global context. To solve this ROI segmentation challenge a generic multi-level multi-tissue segmentation model inspired by Hook-Net~[ref] architecture is proposed which works well for multiple stains across multiple data sources and scanners.  The model’s architecture is shown in Figure~\ref{fig:roi_model}.

As shown in Figure~\ref{fig:roi_model}, the model uses multiple UNet in parallel which enhances the feature capturing capability across resolutions. Our custom Merge Logic uses stepwise 1D convolutions to non-linearly compress the high-dimension features to lower dimensions without concatenation. This helps with the proper context transfer from higher FOV to lower FOV making the model robust with all kinds of tissues. Also, we have combined the output of both the UNet channel-wise by cropping the center of the low-resolution image, resizing it to high-resolution dimensions, and concatenation it with 1D convolutions. This helps the model gain more confidence in the predicted class of the pixel.
We have used efficeintNetB4 as an encoder in both the UNet levels because of the stability and lightweightiness of the model. The output of the model is a 3-channel image-like array with each channel representing one class. We have used random weight initialization with Adam optimizer~[ref] and the equal combination of weighted cross entropy loss and weighted dice loss. Static weights of 1:2 were used for low \& high resolution respectively to control the effect of UNet levels contributing to loss.  The stepwise decremental annealing learning rate starting with 0.001 was used for training the model. 

To take into account the partial, noisy, and smaller annotated region, we have used a 2-phase training approach. In the first phase, using impure datasets, we trained our multi-level model with a small encoder (EfficienetB4) for 20 epochs. This model gave a descent performance of 0.70 IOU but is not generic enough for multiple scanners and data sources. This model is now used to predict the entire dataset including non-annotated slides with a very high threshold of 0.9. This predicted output mask of the first phase becomes input to phase 2 of the training. In Phase 2, we have used our multi-level UNet with a large encoder (EfficeinetB4). This Phase used multiple training steps with increased augmentations and decreased threshold.

In the second Phase, each step of training used slightly different hyperparameters as can be seen from data furnished in Table~\ref{tab:roi_model_hyperparameters}. Static class weights were also used in the last child training to make the model a little biased toward tumor regions. Batch-level pixel-wise dynamic weights are used in the rest of the steps of training.

\begin{table}[!t]
\centering
\resizebox{\textwidth}{!}{%
\begin{tabular}{lccllccccl}
\hline
Step              & \multicolumn{1}{l}{Model} & \multicolumn{1}{l}{Encoder} & Dataset                                                                          & Augmentations                                                                                                                        & \multicolumn{1}{l}{Threshold} & \multicolumn{1}{l}{Epochs} & \multicolumn{1}{l}{LR0} & \multicolumn{1}{l}{Batch Size} & Class Weight \\ \hline
Base Training     & UNet                      & B0                          & \begin{tabular}[c]{@{}l@{}}Manually annotated \\ slides\end{tabular}             & basic - crop, flip, rotate, rescale, normalize                                                                                       & 0.9                           & 20                         & 0.01                    & \multirow{6}{*}{16}            & Static       \\ \cline{1-8} \cline{10-10} 
Master Training-1 & \multirow{5}{*}{m-UNet}   & \multirow{5}{*}{B4}         & \multirow{5}{*}{\begin{tabular}[c]{@{}l@{}}Complete training\\ set\end{tabular}} & \begin{tabular}[c]{@{}l@{}}basic + color jitter \\ + noise addition\end{tabular}                                                     & 0.8                           & 5                          & 0.005                   &                                & Dynamic      \\ \cline{1-1} \cline{5-8} \cline{10-10} 
Master Training-2 &                           &                             &                                                                                  & \begin{tabular}[c]{@{}l@{}}basic \\ + heavy color augmentation\\ + noise addition\end{tabular}                                       & 0.7                           & 5                          & 0.001                   &                                & Dynamic      \\ \cline{1-1} \cline{5-8} \cline{10-10} 
Master Training-3 &                           &                             &                                                                                  & \begin{tabular}[c]{@{}l@{}}Targeted HSV augmentation \\ + basic   \\ + heavy color  augmentation   \\ + noise addition\end{tabular}  & 0.6                           & 5                          & 0.0007                  &                                & Dynamic      \\ \cline{1-1} \cline{5-8} \cline{10-10} 
Master Training-4 &                           &                             &                                                                                  & \begin{tabular}[c]{@{}l@{}}GAN + targeted HSV augmentation\\ + basic  \\ + heavy color augmentation \\ + noise addition\end{tabular} & 0.5                           & 5                          & 0.0005                  &                                & Dynamic      \\ \cline{1-1} \cline{5-8} \cline{10-10} 
Master Training-5 &                           &                             &                                                                                  & \begin{tabular}[c]{@{}l@{}}GAN augmentation + XY coordinate Jitter \\ + basic   \\ + noise  addition\end{tabular}                    & 0.4                           & 10                         & 0.0001                  &                                & Static       \\ \hline
\end{tabular}%
}
\caption{Stepwise hyperparameters such as epochs, learning rate, class weights, etc. that are used for training.}
\label{tab:roi_model_hyperparameters}
\end{table}

\subsubsection*{Nuclei-Detection and Classification}
\begin{figure}[!t]
    \centering
    \includegraphics[width=\textwidth]{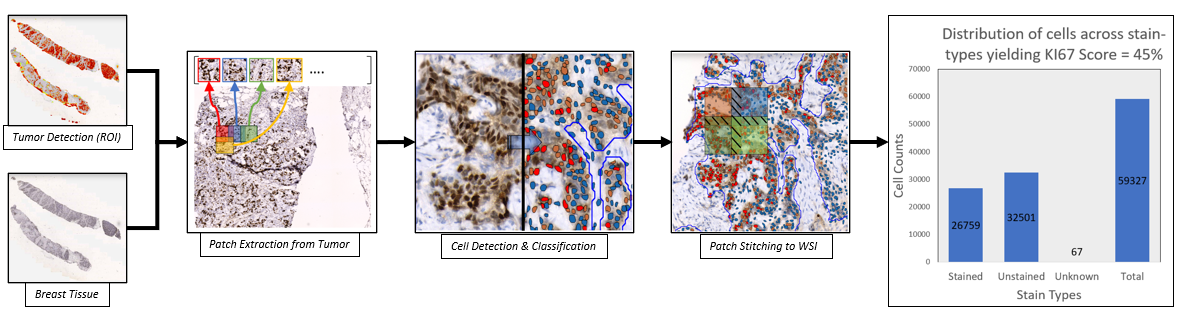}
    \caption{The flow diagram of nuclei segmentation and classification process to compute the Allred score of given digitized breast cancer tissue.}
    \label{fig:nuclei_detect_process}
\end{figure}

The nuclei detection and classification are performed in stage-02 of the DSS. This process is performed for breast tissue stained with ER, PR, and Ki67 IHC markers. The module utilizes the ROI results from stage-01. A brief overview of the complete process can be explained using the flow diagram shown in Figure~\ref{fig:nuclei_detect_process}. In the input to the system is a digitized breast cancer tissue scanned to have a maximum magnification of 40 times. As, the spatial resolution of the digital image can range from mega-pixels to giga-pixels, hence, to reduce the computational complexity of the image patches corresponding to just the tissue region (discarding background or white region in tissue) are extracted. The extracted patches are then used for ROI detection and nuclei segmentation and classification. The nuclei segmentation and classification processes generate results localized to the patch. To create global results, the results of all the patches extracted are stitched together to create a single image. This nuclei segmentation and classification result is then masked with ROI results to limit the process only to the valid tumorous regions in the tissue. Afterward, the obtained results are quantized to compute the Allred score of the tissue as per the CAP (College of American Pathologists) guidelines. 

We used Mask-RCNN [30], which is a CNN-based architecture, for the nuclei segmentation. In the output of the model, the locations of all objects in an image, masks are also generated. Separate instance value is assigned to objects of same class. The architecture has three main parts: 1) The backbone model, 2) the region proposal networks, and 3) Object localization and classification sub-networks. The backbone model is the main feature extractor of the architecture. The ResNet50~\cite{he2016deep} architecture without its classification limb was used as the backbone model in the Mask-RCNN~\cite{he2017mask} architecture.  The feature maps generated at the last layer of ResNet50~\cite{he2016deep} architecture are shared among the region proposal network (RPN) and the object localization and classification sub-networks.
The object’s proposal generated by the RPN layer is then refined by following sub-networks to provide the object’s location information along with the category it belongs. In our use case, the objects category is foreground and background.

The Mask RCNN model is trained by dividing it into three steps based on the layers of the model trained in each of these steps using the Teacher-Student (TS)~\cite{hu2022teacher} training strategy. In the first step, all the trainable parameters of all the layers are trained. The second step trains all the layers beyond the fifth layer of the model while trainable parameters till the fifth layer are frozen. Finally, in the last step, the heads of the model are trained. The head of the model includes all the prediction layers or sub-networks which include the object classification layers, the mask generation layer, and the bounding box regression layer. The training hyperparameters in each of the three steps are different. The epoch values are 5, 5, and 3 for each step, respectively. Similarly, the learning rate values are 1e-4, 1e-5, and 1e-5, respectively. The batch size is a factor of image per GPU and the number of GPUs. We used 2 GPUs with 16 GB VRAM per GPU and a maximum of 4 images per GPU can fit in the given VRAM capacity of the GPUs. The input image size of the RGB patch extracted is $512 \times 512$ which is scaled internally to $768 \times 768$ during the training. A stochastic Gradient Descent (SGD) optimizer with a momentum value of 0.9 and active Nesterov is used to train the model. 

\begin{figure}[!t]
    \centering
     \subfloat[]{\includegraphics[width=0.35\linewidth]{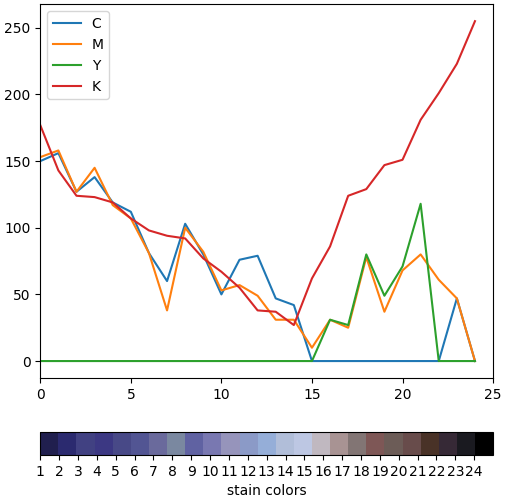}} 
     \subfloat[]{\includegraphics[width=0.35\linewidth]{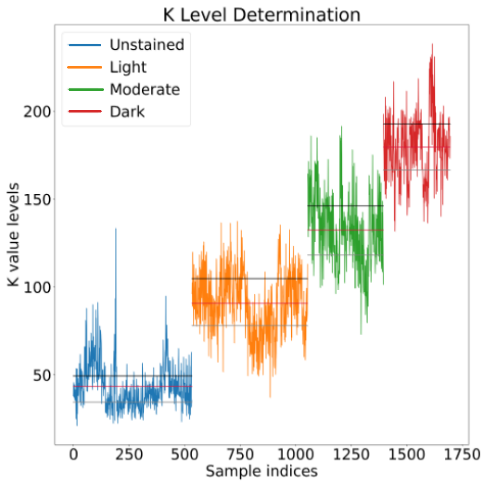}} 
    \caption{(a) The variation in the value of C, M, Y, and K components in CMYK color space for different stain hue and its value on 25 samples extracted from multiple different stained breast tissue WSI. (b) Variation of K-levels across stain types for multiple cells.}
    \label{fig:cmyk_process}
\end{figure}

Following the nuclei segmentation step, the segmentation nuclei are classified into different categories based on the color of their stain. For Ki67, the categories are unstained and stained. For ER and PR, the stain colors are classified into unstained, lightly stained, moderately stained, and dark stained. We used a color-based classification approach. the RGB image of a nuclei is converted to a CMYK color space. Next, using value thresholding on different channels of CMYK color space, the classification of nuclei stain color is performed. While RGB color space is an additive color space, CMYK color space is a subtractive color space where C, M, Y, K means Cyan, Magenta, Yellow, and Black, respectively. As identified by the author [33], there is a relationship between C and Y color values with the unstained and stained nuclei. With the IHC markers, the nuclei take the color in RGB color space of blue and brown called unstained and stain nuclei, respectively. The brownness of nuclei indicates the presence of hormones or proteins while blueness means its absence. In CMYK color space, the value of the Y channel is near zero in the range [0, 255] while the value of the C channel is non-zero (depending on the shade of blue color). Conversely, the C channel is near zero or zero (range is [0, 255]) for brown (stained nuclei) while the Y value is non-zero. The effect of unstained and stained nuclei color on the values of different components of CMYK color space is shown in Figure~\ref{fig:cmyk_process}.

Additionally, it is observed that the K channel in CMYK color space has an approximately linear relation with the intensity of unstained or stain color. The same can be seen in Figure 10a. The color samples shown in the palette are taken from actual nuclei present in the WSI. The red color curve in Figure 10a shows the variation in the value of the K color component with changing color intensity of either unstained or stained nuclei.  Figure 10b shows the variation value of K channels for unstained, lightly stained, moderately stained, and dark-stained nuclei categories. The separation between different categories is evident in this graph. Hence, using different channel pixel values the nuclei are categorized. The segregation of unstained and stained nuclei is done using the C and Y color components of the CMYK color space. Moreover, the K color component is used for the classification of stained nuclei into light, moderate, and dark categories. We used value thresholding to perform the same. In total three different values are used namely $\delta_u$, $\delta_{sl}$, and $\delta_{su}$. Here, $\delta_{u}$ is the threshold value applied to the Y component to separate unstained and stained nuclei. $\delta_{sl}$ and $\delta_{su}$ are applied on the K color component to classify stained cells into light (L), moderate (M), and dark (D) categories. 

Mathematically,
\begin{align}
    SNL = \begin{cases}
        lightly stained, & if K < \delta_{sl},\\
        dark stained, &if K > \delta_{su}, \\
        moderately stained, & otherwise
    \end{cases}
\end{align}
Here, it should be noted that $\delta_{sl} < \delta_{su}$ and SNL is to refer stained nuclei label.

The values of these thresholds are computed using the pixel values of tumorous nuclei from the annotated ground-truth data.

\subsubsection*{Membrane-based tissue analysis}
\begin{figure}[!t]
    \centering
    \includegraphics[width=\textwidth]{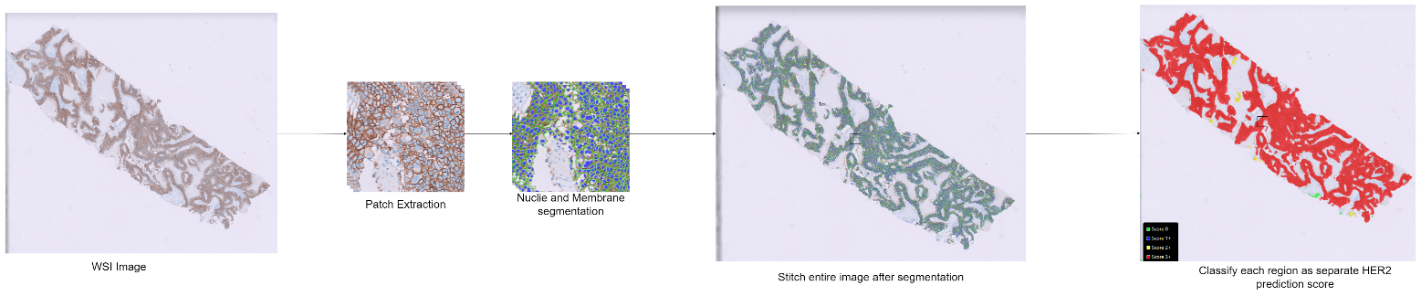}
    \caption{The proposed approach to solve the HER2 membrane as two-step model.}
    \label{fig:membrance_score_process}
\end{figure}

All breast cells have and are tested for an excess of human epidermal growth factor receptor 2, commonly referred to as HER2. HER2 proteins are receptors that control how the cells grow and divide. The immunostaining results of HER2 are reported as follows: 0 as negative; 1+ as negative; 2+ as equivocal; and 3+ as positive~\cite{fauzi2022allred}. These scores are computed based on the connectivity and color profile of membranes in the breast tissue. In our DSS, to compute the HER2 score, we used deep learning and machine learning. The first step is to mimic the working of the pathologist in understanding where the membrane and nuclei are present in the entire slide, this is done by the first deep learning-based model, the second step after this is to classify the various regions in the slide HER2 scores and also give out overall slide level prediction by combining all regions as 1 single region, This is carried out by the second machine learning model. The overall approach can be understood through the process delineated in Figure~\ref{fig:membrance_score_process}.

\textbf{\textit{First model}}: Membrane and Nuclei segmentation: The membrane and nuclei segmentation model gets an RGB image patch. The patches are extracted from level 1 of the pyramidal image at $20x$ magnification. These image patches are fed to a CNN-based model which does semantic segmentation of the membrane lining and nuclei detection. We used EfficientNet-UNET, which is a CNN-based architecture, for the semantic segmentation. The EfficientNet-B4 UNet model for semantic segmentation in HER2 image patches combines the EfficientNet-B4~\cite{tan2019efficientnet} architecture with the UNet architecture. 

In this context, the model aims to segment membrane and nuclei in HER2 image patches. The EfficientNet-B4~\cite{tan2019efficientnet} backbone efficiently extracts hierarchical features from the input images, capturing both low-level and high-level representations. The UNet architecture facilitates precise segmentation by combining the high-resolution features from the contracting path with upsampled coarse features from the expansive path. Key components include the contracting path, which captures contextual information, and the expansive path, which refines segmentation details. Skip connections between corresponding layers in the contracting and expansive paths help preserve spatial information, enhancing the model's ability to delineate membrane and nuclei boundaries. The combination of EfficientNet-B4~\cite{tan2019efficientnet} and UNet~\cite{ronneberger2015u} addresses the challenge of balancing efficiency and accuracy, making it suitable for processing HER2 image patches. Experimentation with loss functions, optimization techniques, and appropriate data augmentation strategies further refine the model's performance in accurately segmenting membrane and nuclei structures in HER2 images. 

In the above model, to overcome the challenges faced due to the nature of the HER2 membrane and nuclei morphology, which are variations in color and width of the membrane and visibility of nuclei in the image due to overstaining, we introduce more non-linear activation-based depth in the UNet model~\cite{ronneberger2015u} to counter that. This is done for the model to have a more granular understanding of the features and is more generic.

The training data for the model is 152 numbers of manually annotated patches from two experts pathologists after the first model is finalized, the second model is trained using the prediction from the first model in a cyclic manner. In this case, six cycles of training are performed. The data used is checked using heuristic logic whether that needs to be retained for training in further cycles or not. In each cycle, there are some hard negatives introduced for the model as well. The final cycle is much smaller and taken from the original model.

\begin{table}[]
\centering
\begin{tabular}{c|c|l}
\multicolumn{1}{l}{Data Source} & WSI & Patches          \\ \hline
Source-01                       & 100 & Cycle 1: 10,000  \\ \hline
\multirow{5}{*}{Source-02}      & 60  & Cycle 2: 30,000  \\
                                &     & Cycle 3: 90,000  \\
                                &     & Cycle 4: 90,000  \\
                                &     & Cycle 5: 130,000 \\
                                &     & Cycle 6: 9,000   \\ \hline
\end{tabular}
\caption{Training data count used for membrane detection and classification}
\label{tab:membrane_train_data}
\end{table}

Now, with this model, a lot of experiments are conducted to maximize the IOU for membrane detection and nuclei detection. The intersection over union (IoU) values for membrane- and nucleus-detection obtained are 38.75\% and 57.26\%, respectively.

\begin{figure}[!ht]
    \centering
    \includegraphics[width=0.5\textwidth]{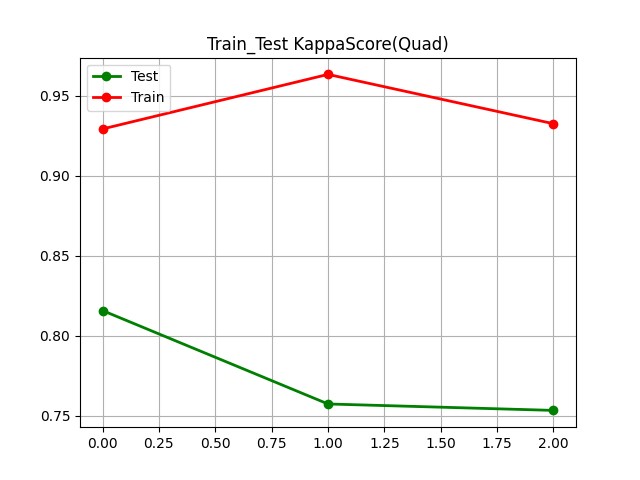}
    \caption{Test train data kappa scores (Quadratic).}
    \label{fig:membrance_kappa_score}
\end{figure}

\textbf{\textit{Second Model}}: Feature to HER2 score Classification: During the stitching process of patches after a prediction from the first model, we calculate some morphological features at each ROI level and for entire slide such as color histogram under the membrane and nuclei, skewness of the histogram curve, nuclei to membrane ratio, percentage cells with complete membrane, etc. These features for the 200 WSI images are computed and stored. Post this the Random Forest model is trained on this 100 WSI data and validated against of other 100 WSI images to predict the HER2 scores for each region of ROI as well as the entire slide. The entire process is performed using K-fold cross-validation with Kappa Score (Quadratic) Figure~\ref{fig:membrance_kappa_score}.

\bibliography{ihc_paper_reference}

\section*{Acknowledgements}
We would like to thank the management at Applied Materials for supporting our work in this domain. We are deeply grateful to the pathologists who provided IHC scores annotations for this work, including Dr. Kanthilatha Pai, Dr. Brij Mohan Kumar Singh, Dr. Anil Betigeri, Dr. Vani Verma, and Dr. Madhavi Pai. We would also like to extend our gratitude to all the domain experts involved in the knowledge sharing, slide quality control, and output validation.

\section*{Author contributions}
S.J., A.M., G.S., S.M., and R.K. were involved in the planning and designing of the experiments. S.J., and K.A., were involved in acquiring the data to perform the experiments and provide strategic support. S.J. and A.M. wrote the code to achieve different tasks. G.S. and S.M. were involved in validating the annotation of the data used during the training process. The results collection and performance analysis was conducted by A.M., and R.K.. S.J., A.M., and P.M. wrote the manuscripts with the assistance and feedback of all other co-authors.

\section*{Competing interests}
The author declares no competing interests.

\section*{Data Availability}
The datasets used in this study were collected as per an internal agreement between Applied Materials India Pvt. Ltd. and hospitals/laboratories participating in this study. The slides and doctor's annotations are not available publicly due to restrictions in the data-sharing agreements with the participating institutions. Data are however available from the authors upon reasonable requests and with the permission of competent authority at Applied Materials India Pvt. Ltd. and participating institutions.




\end{document}